\documentclass[12pt]{article}
\usepackage{cite}
\usepackage{color}
\usepackage{graphicx}

\makeatletter
\@addtoreset{equation}{section}

\makeatletter
\renewcommand\section{\@startsection {section}{1}{\z@}%
                                   {-3.5ex \@plus -1ex \@minus -.2ex}
                                   {2.3ex \@plus.2ex}%
                                   {\normalfont\large\bfseries}}
\renewcommand\subsection{\@startsection{subsection}{2}{\z@}%
                                     {-3.25ex\@plus -1ex \@minus -.2ex}%
                                     {1.5ex \@plus .2ex}%
                                     {\normalfont\bfseries}}

\def\baselinestretch{1.2}
\newcommand{\be}{\begin{equation}}
\newcommand{\ee}{\end{equation}}
\newcommand{\beq}{\begin{eqnarray}}
\newcommand{\eeq}{\end{eqnarray}}
\newcommand{\tr}{{\rm tr}}

\begin{document}
\begin{titlepage}
\begin{flushright}
hep-th/0511197\\
MAD-TH-05-07
\end{flushright}

\vfil\

\begin{center}

{\large{\bf Non-commutative gauge theory on D-branes in Melvin Universes}}

\vfil

Akikazu Hashimoto and Keith Thomas

\vfil

Department of Physics\\
University of Wisconsin\\ Madison, WI 53706\\

\vfil

\end{center}

\begin{abstract}
\noindent Non-commutative gauge theory with a non-constant
non-commutativity parameter can be formulated as a decoupling limit of
open strings ending on D3-branes wrapping a Melvin universe.  We
construct the action explicitly and discuss various physical features
of this theory.  The decoupled field theory is not supersymmetric.
Nonetheless, the Coulomb branch appears to remain flat at least in the
large $N$ and large 't Hooft coupling limit. We also find the analogue
of Prasad-Sommerfield monopoles whose size scales with the
non-commutativity parameter and is therefore position dependent.
\end{abstract}
\vspace{0.5in}

\end{titlepage}
\renewcommand{\baselinestretch}{1.05}  

\section{Introduction}

It is generally believed that a proper understanding of quantum
gravity requires taking quantum field theory beyond a framework based
on locality.  Non-commutative geometry is an important conceptual
laboratory for exploring precisely this issue.

Gauge theories on non-commutative spaces arise naturally as a certain
decoupling limit of open string dynamics in the presence of a
background NSNS $B$-field.  In recent years, there has been
significant progress in our understanding of non-commutative field
theories on flat spaces with constant non-commutativity.  These
theories arise in cases where the $B$-field is uniform
\cite{Douglas:2001ba}. Less is known about the generalization to the
case where the non-commutativity parameter is non-constant. See
e.g.\cite{Madore:2000en,Ho:2000fv,Cornalba:2001sm,Ho:2001fi,Herbst:2001ai,Cerchiai:2003yu,Calmet:2003jv,Bertolami:2003nm,Robbins:2003ry,Behr:2003qc,Das:2003kw,Gayral:2005ih}.
Nonetheless, several concrete examples of non-commutative gauge
theories with non-constant non-commutativities have been shown to
arise as a decoupling limit of open strings in specific backgrounds
with nontrivial background NSNS $B$-field
\cite{Dolan:2002px,Hashimoto:2002nr,Lowe:2003qy,Hashimoto:2004pb}.

A large class of backgrounds that can give rise to non-commutative
gauge theories in the decoupling limit can be constructed by acting on
flat space with twists and dualities. A particularly simple example of
this construction is the Melvin universe \cite{Melvin:1963qx}
supported by the field strength of the NSNS $B$-field
\cite{Russo:1995tj}. The decoupling limit of open string dynamics in
this background was considered in \cite{Hashimoto:2004pb}.  It is a
non-commutative gauge theory with the matter content of ${\cal N}=4$
supersymmetry and a non-commutativity parameter that is position
dependent.  Such a theory can be formulated in terms of an action
where the fields are multiplied using the $*$-product of Kontsevich
\cite{Kontsevich:1997vb}. The goal of this paper is to explore the
physical features of this model in some detail.

The structure of supersymmetry is intricate for this theory. It is
well known that type II string theory in Melvin universes generically
breaks all of the supersymmetries
\cite{Russo:1995ik,Costa:2000nw,Gutperle:2001mb}; the spectrum of open
string fluctuations respects the Bose-Fermi degeneracy
\cite{Takayanagi:2001aj} but the supersymmetries are broken by the
interactions.  An easy way to see that this theory is not
supersymmetric is to note that translation invariance is broken due to
the non-constant non-commutativity parameter.  This is incompatible
with the supersymmetry algebra.  However, we also note that
supersymmetry is nonetheless restored in the infrared where the
effects of non-commutativity becomes irrelevant.

In the large $N$ and large 't Hooft coupling limit, it is straight
forward to compute the DBI action of a probe D3-brane which measures
the possible lifting of the Coulomb branch due to quantum effects
\cite{Maldacena:1997re}. It turns out however that no potential is
generated along the Coulomb branch in the regime where the
supergravity analysis is reliable.  This is somewhat surprising in
light of the fact that supersymmetry is clearly broken.  It appears
that the effects of supersymmetry breaking are very mild for this
theory.

Since the Coulomb branch is flat, this gauge theory can support
Prasad-Sommerfield monopoles \cite{Prasad:1975kr} in its broken
phase. These monopoles have sizes proportional to the value of the
non-commutativity parameter \cite{Hashimoto:1999zw,Gross:2000ss} and
as a result their size is position dependent in our model. They
therefore provide a convenient probe for measuring the local value of
the non-commutativity parameter.

This paper is organized as follows. In section 2, we review the
construction of the Melvin universe in the presence of D3-branes as a
background of type IIB string theory and describe the decoupling limit
which gives rise to a non-commutative gauge theory with non-constant
non-commutativity. In section 3, we present an explicit expression for
the action for this gauge theory and analyze it's supersymmetry. In
section 4, we construct the dual supergravity description for this
theory and study the killing symmetry of that background. We also
describe the Coulomb branch and the Prasad-Sommerfield monopoles of
this model. We conclude in section 5.
\section{IIB Melvin Universes and Non-commutative Gauge Theory}

The Melvin universe is a non-asymptotically flat solution to the type
IIB supergravity equations of motion.  It has topology $R^3 \times R^1
\times R^6$ and is supported by the flux of a space-time dependent
NSNS $B$-field.  This background can be constructed by acting on flat
space with a chain of twists and T-dualities.  Upon wrapping D3-branes
around this geometry and taking a specific decoupling limit, one
obtains a non-commutative gauge theory with a space-time dependent
non-commutativity parameter whose magnitude is proportional to the
twist parameter.  In this section we review the construction of the
Melvin Universe and the decoupling limit.

The Melvin Universe can be obtained by applying the following chain of
operations to flat space.
\begin{enumerate}
\item Start with a flat background in type IIB supergravity
\be ds^2 = -dt^2 + dr^2 + r^2 d \varphi^2 + dz^2 + \sum_{i=1,6} dy_i^2 \ee
where $z$ is compactified on a circle with radius $R$.
\item T-dualize along $z$ to obtain a background of type IIA supergravity
\be ds^2 = -dt^2 + dr^2 + r^2 d \varphi^2 + d \tilde z^2 + \sum_{i=1,6} dy_i^2 \ee
where the radius of $\tilde z$ coordinate is $\tilde R = \alpha' / R$.
\item This geometry admits an isometry generated by a vector
${\partial / \partial \varphi}$. Given such an isometry vector, one
can ``twist'' the compactification. By this, one means changing the
Killing vector associated with the compactification from $({\partial /
\partial \tilde z})$ to $({\partial / \partial \tilde z} + \eta {
\partial / \partial \varphi})$.  Alternatively, one can think of the
twist as first replacing
\be d\varphi  \rightarrow d \varphi + \eta \, d \tilde z \ee
so that the metric reads
\be ds^2 = -dt^2 + dr^2 + r^2 (d \varphi + \eta d\tilde z)^2 + d \tilde z^2 + \sum_{i=1,6} dy_i^2 \label{IIAD2}\ee
and then treating $\tilde z$ as the periodic variable of radius
$\tilde R$ with $\varphi$ fixed.
\item T-dualize along $\tilde z$ to obtain a solution of type IIB supergravity
\beq
ds^2  & = &  -dt^2 + dr^2 + {r^2 \over 1 + \eta^2 r^2} d\varphi^2 +  {1  \over 1 + \eta^2 r^2} d  z^2 + \sum_{i = 1}^6 dy_i^2  \cr
B & = & {\eta r^2 \over 1+\eta^2 r^2} d \varphi \wedge d  z  \label{melvin} \\
e^\phi & = & g_s \sqrt{{1 \over 1 + \eta^2 r^2}} \ . \nonumber
\eeq
\end{enumerate}
This is a Melvin Universe supported by the flux of the NSNS $B$-field.
It's global geometry is that of a teardrop.

In order to obtain a gauge theory, we consider D3-branes extended
along the $\left(t,r,\varphi,z \right)$ coordinates and apply the
mapping of Seiberg and Witten \cite{Seiberg:1999vs}
\be (G + {\theta \over 2 \pi \alpha'})^{\mu \nu} = [(g + B)_{\mu \nu}]^{-1} \ .  \label{swmap} \ee
Applying this to the closed string background (2.5) we obtain the open
string metric $G$ and non-commutativity parameter $\theta$
\beq G_{\mu \nu} dx^\mu dx^\nu &=& -dt^2 + dr^2 + r^2 d \varphi^2 + d  z^2 \cr
\theta^{\varphi  z} & = & 2 \pi \alpha' \eta \ . \label{openmetric}
\eeq
The metric (2.7) is independent of $\alpha'$ and will therefore be the
background for the decoupled theory.  In order for the field theory on
this background to have finite non-commutativity, we must take the
particular scaling limit
\be
  \eta = \frac{\Delta}{\alpha'}  \hspace{8mm} \alpha' \rightarrow 0 \ .
\ee
With this choice of scaling, the non-commutativity parameter has a
finite limit
\be \theta^{\varphi z} = - \theta^{z \varphi} = 2 \pi \Delta  \label{polartheta} \ . \ee 
Further, it is clear in cartesian coordinates that the
non-commutativity parameter is non-constant
\be \theta^{x_1 z} = -\theta^{z x_1}  = -2 \pi \Delta x_2, \qquad \theta^{x_2 z} =-\theta^{z x_2} = 2 \pi \Delta x_1 \label{melvintheta} \ .\ee
It is divergence free
\be \partial_i \theta^{ij} = 0 \ . \ee
and satisfies the  Jacobi identity 
\be \theta^{il} \partial_l \theta^{jk}
+ \theta^{jl} \partial_l \theta^{ki}
+ \theta^{kl} \partial_l \theta^{ij} = 0 \ee
defining a Poisson structure. These features are useful in
constructing a gauge invarant action on non-commutative spact-times
\cite{Behr:2003qc,Calmet:2003jv,Felder:2000nc}.

\section{Non-commutative gauge theory as a D-brane world volume theory}

In this section, we write down the action for the non-commutative
gauge theory we obtained in the decoupling limit of D3-branes embedded
in the Melvin geometry (\ref{melvin}).  Because the non-commutativity
tensor $\theta^{\mu \nu}$ we found in (\ref{melvintheta}) is
non-constant, we construct an associative product using the formula of
Kontsevich \cite{Kontsevich:1997vb}
\beq f* g &=& fg + i {\theta^{\mu \nu} \over 2} \partial_\mu f \partial_\nu g   - {1 \over 8} \theta^{\mu \nu} \theta^{\lambda \sigma} \partial_\mu \partial_\lambda f \partial_\nu \partial_\sigma g \cr
&& - {1 \over 12} \theta^{\mu  \nu} \partial_\nu \theta^{\lambda \sigma}
\left(\partial_\mu \partial_\lambda f \partial_\sigma g 
- \partial_\lambda f \partial_\mu \partial_\sigma g \right) + {\cal O}(\theta^3) \ . \eeq 
Taking $\theta$ constant in this formula gives the Moyal product.
Naively, the action for non-commutative gauge theory is obtained by
beginning with the ordinary gauge theory action and replacing all
multiplications of the fields by the $*$-product.

This however is not suitable for constructing a gauge invariant action
when the non-commutativity parameter is position dependent;
differentiation does not respect the product rule
\be \partial_\mu (f*g) \ne \partial_\mu f * g + f * \partial_\mu g \ . \ee 
In order to properly formulate the action of non-commutative gauge
theory, one can use the frame formalism introduced in
\cite{Behr:2003qc}. This procedure will work for a generic Poisson
bivector $\theta^{\mu \nu}$ on an arbitrary curved manifold. The
non-commutative field theory we consider is however extremely simple
and one can write down an explicit form of the action in a rather
straight forward manor. The result is in agreement with the general
treatment of \cite{Behr:2003qc}.

To do this, first recall from the previous section that the
non-commutativity parameter takes on a simple constant form in polar
coordinates (\ref{polartheta}).  One can therefore define a Moyal-like
product
\be f \# g = \left. e^{{i \theta^{\varphi z} \over 2} (
\partial_\varphi \partial_{z'} - \partial_{\varphi'} \partial_z)}
f(t,r,\varphi,z) g(t,r,\varphi',z') \right|_{{\varphi=\varphi'\atop z=z'}} \ .  \ee
The $\#$-product and $*$-product are not simply related by a change of
coordinates, but are equivalent in the sense defined by Kontsevich
\cite{Kontsevich:1997vb}
\be
R(f \# g) = R(f) * R(g) \ .
\ee
To order ${\cal O}(\theta^2) \sim {\cal O}(\Delta^2)$, $R(f)$ is given
by
\be
 R(f) = f + \frac{4 \pi^2 \Delta^2}{24} r \partial_{r} \partial_z^2 R + {\cal  O}(\Delta^3) \ .
\ee
Similar analysis of an explicit form of $R(f)$ can be found in
\cite{Cerchiai:2003yu}.

In polar coordinates, it is natural to define the following set of
unit norm vector fields
\be \partial_1 = \partial_t, \qquad \partial_2 = \partial_r, \qquad \partial_3 = {1 \over r} \partial_\varphi, \qquad \partial_4  = \partial_z \ . \ee
They can also be written in terms of components
\be \partial_a = X_a^\mu \partial_\mu \ . \ee
The vectors $X_a$ define a natural local frame and one can therefore
write the action of ordinary non-abelian gauge theory in the form
\be S = {1 \over 2} \tr  \int \sqrt{-G} \,  G^{ab} G^{cd} \, F_{ac} F_{bd} \ee
where $G_{ab} = g_{\mu \nu} X_a^\mu X_b^\nu = \eta_{ab}$ is the flat
metric on the local frame, and
\be F_{ab} = \partial_a A_b - \partial_b A_a + i g[ A_a, A_b]\,
,\qquad A_a = X_a^\mu A_\mu \ .  \ee
In this formalism, it is straight forward to sprinkle the $\#$'s to
define a non-commutative theory
\be S = {1 \over 2} \tr  \int \sqrt{-G} \,  G^{ab} G^{cd} \, F_{ac} \# F_{bd},
\qquad
F_{ab}  =  \partial_a A_b - \partial_b A_a + i g A_a \#  A_b - i g A_b \#  A_a \ee
since the $\partial_a$'s  respect the product rule
\be \partial_a (f \# g) =  \partial_a f \# g +  f \# \partial_a  g \ . \ee 
However, some care is required in working with the action in this
form.  In particular, as integration by parts does not apply for one
of the derivatives
\be \partial_3 = \partial_r \ee
particular care is required when deriving the equations of motion.

Using these results and the automorphism $R(f)$, it is straight
forward to formulate the action using the $*$-product.  A derivation
respecting the product rule can be defined according to
\be \delta_{X_a} f = R \partial_a R^{-1}  f \ee
since
\be \delta_{X_a} (f * g) = \delta_{X_a} R  (R^{-1} f \# R^{-1} g) = 
\delta_{X_a} f * g + f * \delta_{X_a}  g  \ .\ee
Then, the  action 
\be S = {1 \over 2} \tr  \int \sqrt{-G} \,  G^{ab} G^{cd} \, F_{ac} * F_{bd},
\qquad
F_{ab}  =  \delta_{X_a} A_b - \delta_{X_b} A_a + i g A_a *  A_b - i g A_b*  A_a  \ee
takes the form in which it was given in \cite{Behr:2003qc}. It is
straight forward to generalize this construction to scalar and spinor
fields and the non-commutative action for a field theory with ${\cal
N}=4$ matter content is given by
\beq
\nonumber S &=& \mbox{tr} \int d^4 x \sqrt{-G} \left[ - \frac{1}{2} G^{ab}G^{cd} F_{ac} * F_{bd} - G^{ab} \sum_i D_a \Phi^i * D_b \Phi^i  \rule{0ex}{4ex} \right. \\
 && \hspace{28mm} \left.  - i   \bar{\psi} * \Gamma^{\mu} X_{\mu}^a D_a \psi+ \frac{1}{2} \sum_{ij}[\Phi^i,\Phi^j]_{*}^2 \hspace{2mm} +\ldots \right] 
\eeq
where 
\be
D_a \psi = \delta_{X_a} \psi - ig \left[ A_a , \psi \right]_{*}
\ee
is the covariant derivative, and the ellipsis indicates the
interaction terms involving the fermions.

In the remainder of this paper, we will explore various physical
features of this theory. One issue which can be addressed immediately
is that of supersymmetry.  The matter content of this theory is
consistent with ${\cal N}=4$ supersymmetry. However, the fact that the
non-commutativity parameter is not translation invariant implies that
supersymmetry (which closes to translation) must be broken. This can
be confirmed by explicit variation of the action. On the other hand,
in the low energy limit where the effect of non-commutativity becomes
irrelevant, all 32 supersymmetries of the ${\cal N}=4$ theory are
restored.  Supersymmetry is also restored in the non-interacting limit
$g_{YM}^2 \rightarrow 0$, since non-commutativity only affects the
interactions.  This is consistent with the comment in footnote 36 of
\cite{Takayanagi:2001aj} concerning supersymmetry of the open string
spectrum for the related embedding of D-branes in the Melvin universe.

\section{Supergravity dual description of the non-commutative gauge theory}

An effective approach in exploring the physical properties of a field
theory is to study its supergravity dual.  In this section, we review
the supergravity dual and analyze three features of our model:
supersymmetry, the Coulomb branch, and magnetic monopoles.

The supergravity background dual to the non-commutative gauge theory
of interests was found in \cite{Hashimoto:2004pb}. The background can
be derived by beginning with the supergravity solution for a stack of
D3-branes
\beq ds^2 &=& f(\rho)^{-1/2}(-dt^2 + dr^2 + r^2 d \varphi^2 + dz^2) + f(\rho)^{1/2}(d\rho^2 + \rho^2 d \Omega_5^2), \cr
f(\rho) &=& 1 + {4 \pi g N \alpha'^2  \over \rho^4} \eeq
and following the chain of dualities outlined in section 2.  Then
perform the scaling
\be \rho = \alpha' U, \qquad \eta = {\Delta \over \alpha'}  \label{horison} \ee
while sending $\alpha' \rightarrow 0$ and keeping $U$ and $\Delta$
fixed.  The resulting geometry in the string frame is
\beq
\nonumber ds^2 &=& \alpha' \left( \frac{U^2}{\sqrt{\lambda}} \left[ -dt^2 + dr^2 + \frac{r^2 d\varphi^2 + dz^2}{1 + {\Delta^2 r^2 U^4 \over \lambda}} \right] + \frac{\sqrt{\lambda}}{U^2} \left[dU^2 + U^2 d\Omega_5^2 \right] \right) \\
\nonumber \\
\nonumber B_{\varphi z} &=& \alpha' \frac{\Delta U^4 r^2}{\lambda +  \Delta^2 U^4 r^2} \hspace{21mm} C_{tr} =  \alpha'  \frac{2 \pi}{g_{YM}^2} \frac{\Delta U^4 r}{\lambda}  \\
\nonumber \\
C_{tr \varphi z} &=& \alpha'^2  \frac{2 \pi}{g_{YM}^2} \frac{U^4 r}{\lambda + \Delta^2 U^4 r^2} \hspace{12mm}        e^{\phi} = \frac{g_{YM}^2}{2 \pi} \sqrt{\frac{\lambda}{\lambda +  \Delta^2 U^4 r^2}} \label{dual}
\eeq
where $ \lambda = 2 g_{YM}^2 N $ .  

Supersymmetry for this solution can be analyzed along the lines of
\cite{Nishimura:2003jd}.  In particular, we obtain the killing spinor
equations for the Melvin Twisted D3-brane background and then consider
the near horizon limit (\ref{horison}).  Using a set of canonical
vielbeins
\beq
E_{t}^0 &=& E_{r}^1 = f^{-1/4}   \hspace{10mm} \frac{E_{\varphi}^2}{r} = E_z^3 = \frac{f^{1/4}}{\sqrt{1+\eta^2 r^2}} \hspace{10mm} E_{\rho}^4 = f^{1/4} \cr
E_{\varphi_5}^5 &=& \rho f^{1/4} \hspace{10mm} E_{\varphi_6}^6 = \rho f^{1/4} \sin{\varphi_5} \hspace{4mm}....\hspace{4mm} E_{\varphi_9}^9 = \rho f^{1/4} \prod_{j=5..8} \sin{\varphi_j}
\eeq
with all other components vanishing, it is straight forward to obtain
the dilatino variation
\be
-\frac{1}{2}e^{-\frac{1}{4} \phi} \hat{\Gamma}^{\mu} \epsilon^* \partial_{\mu} \phi + \frac{1}{2} e^{\phi} \hat{\Gamma}^{\mu \nu \rho} \epsilon {\cal G}_{\mu \nu \rho} \ .
\ee
Here $\hat{\Gamma}^{\rho} = E^{\rho}_a \Gamma^a$ with the $\Gamma^a$
being the standard $32 \times 32$ gamma matrices of Minkowski
space-time in $10$ dimensions, and
\be
{\cal G}_{\mu \nu \rho} = i \left[ 3 \partial_{[\mu} C_{\nu \rho] } + 3ie^{-\phi}  \partial_{[\mu} B_{\nu \rho] } \right] \ .
\ee
The corresponding killing spinor equations are
\be
\eta \left( \Gamma^{123} \epsilon + \eta r \Gamma^1 \epsilon^* \right) = 0  \label{ks}
\ee
where $\epsilon$ is a chiral spinor satisfying $\Gamma^{10} \epsilon =
-\epsilon$ where $\Gamma^{10} = \Gamma^0 \Gamma^1 \ldots
\Gamma^9$. There is no non-tirvial solution to (\ref{ks}) either
before or after taking the near horizon limit.  This shows that as
expected from the field theory analysis, supersymmetry is broken in
the supergravity dual.

\begin{figure}
\centerline{\begin{picture}(0,0)%
\includegraphics{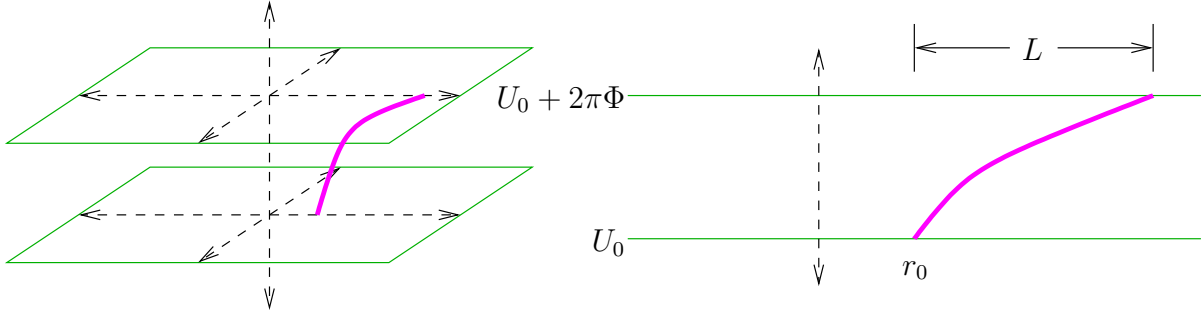}%
\end{picture}%
\setlength{\unitlength}{3947sp}%
\begingroup\makeatletter\ifx\SetFigFont\undefined%
\gdef\SetFigFont#1#2#3#4#5{%
  \reset@font\fontsize{#1}{#2pt}%
  \fontfamily{#3}\fontseries{#4}\fontshape{#5}%
  \selectfont}%
\fi\endgroup%
\begin{picture}(7524,1974)(1489,-2323)
\put(5176,-1036){\makebox(0,0)[lb]{\smash{\SetFigFont{12}{14.4}{\rmdefault}{\mddefault}{\updefault}{\color[rgb]{0,0,0}$\!\!\!\!\!\!\!\!\!\!\!\!\!\!\!\!\!\!U_0+2 \pi \Phi$}%
}}}
\put(7126,-2086){\makebox(0,0)[lb]{\smash{\SetFigFont{12}{14.4}{\rmdefault}{\mddefault}{\updefault}{\color[rgb]{0,0,0}$r_0$}%
}}}
\put(5176,-1936){\makebox(0,0)[lb]{\smash{\SetFigFont{12}{14.4}{\rmdefault}{\mddefault}{\updefault}{\color[rgb]{0,0,0}$U_0$}%
}}}
\put(7876,-736){\makebox(0,0)[lb]{\smash{\SetFigFont{12}{14.4}{\rmdefault}{\mddefault}{\updefault}{\color[rgb]{0,0,0}$L$}%
}}}
\end{picture}
}
\caption{D-brane construction of the position dependent monopole}
\end{figure} 
A simple physical observable of the field theory which can be computed
from the supergravity dual is the potential along the Coulomb branch.
Classically, these field theories have flat directions along field
directions for which $[\Phi^i,\Phi^j] = 0$.  For the ${\cal N} = 4$
SYM theory at strong coupling, the potential along the classical flat
direction was computed using the DBI action of a probe D3-brane in
$AdS_5 \times S_5$ \cite{Maldacena:1997re}.  It is straight forward to
repeat this analysis for our model.  Consider the DBI action for a
probe D3-brane localized in $U$ and extended along $(t,r,\varphi,z)$
of the Melvin-Twisted D3-brane background
\vspace{2mm}
\be
S = - \frac{1}{8 \pi^3 \alpha'^2} \int_M d^4 x\,  e^{-\phi} \det{}^{1/2} {\left[G_{ab} + B_{ab} \right]}+ \frac{1}{8 \pi^3 \alpha'^2} \int_M \left(C^{(4)} + C^{(2)} \wedge B \right) \ .
\ee
When the background (\ref{dual}) is substituted into this action, one
finds
\be
S = -\frac{1}{8 \pi^3\alpha'^2}\left[ \int d^4 x \,  \frac{U^4}{\sqrt{\lambda}} - \int d^4 x\,  \frac{U^4}{\sqrt{\lambda}}  \right] \ .
\ee
The action on this background is therefore $\Delta$-independent and
the NSNS and RR contributions cancel identically just as in the
$\Delta = 0$ case \cite{Maldacena:1997re}.  This implies that the
Coulomb branch remains flat for the non-commutative theory at large 't
Hooft coupling, even when $U^4 \gg \frac{\lambda}{\Delta^2 r^2}$ where
the deformation of the supergravity background (\ref{dual}) from $AdS
_5\times S_5$ is large. This is somewhat surprising as one would
expect the moduli space to receive corrections.  One possibility is
that the moduli space being a zero momentum observable is insensitive
to the non-commutativity.  Nevertheless, it is surprising that there
does not seem to be contributions coming from loop corrections.  In
regard to the potential along the Coulomb branch, it therefore appears
that effect of supersymmetry breaking is mild.

The fact that the Coulomb branch is flat implies that one can turn on
vacuum expectation values for the adjoint scalars and support a
Prasad-Sommerfield monopole \cite{Prasad:1975kr}.  Simple physical
properties of these monopoles can then be studied quite effectively
using the dual supergravity formalism \cite{Hashimoto:1999zw}.  In
this spirit, consider a probe D-string suspended between a pair of
D3-branes in the background (\ref{dual}).  Let the the pair of
D3-branes be located at $U=U_0$ and $U = U_0+2 \pi \Phi$.  The DBI
action for the probe D-string is given by
\be 
S = -\frac{1}{2 \pi \alpha'} \int_{U_0}^{U_0 + 2 \pi \Phi} dU  \left[ e^{-\phi} \sqrt{ -G_{tt} \left(G_{UU} + r'(U)^2 G_{rr} \right)} - C_{0r} r'(U) \right] \label{Dstringaction}
\ee
where we have parameterized the static configuration of the string by
a function $r = r(U)$.  When this ansatz and the background
(\ref{dual}) are substituted into (\ref{Dstringaction}) the action
becomes
\be S = - {1  \over g_{YM}^2 \lambda}  \int_{U_0}^{U_0+2 \pi \Phi} dU \left[ 
   {\sqrt{\left( \lambda + {\Delta}^2 U^4 {r(U)}^2 \right)  
       \left( \lambda + U^4 {r'(U)}^2 \right) }}
 - \Delta U^4 r(U) r'(U) 
\right]  \ . \label{explicit} \ee
The boundary conditions on the $D$-string imply that
\be
\frac{dr}{dU} = \Delta r  \label{bc}
\ee
at the boundaries $U=U_0$ and $U = U_0 + 2 \pi \Phi$. The solution to
(\ref{bc})
\be
U = U_0 + \frac{1}{\Delta} \ln{\frac{r}{r_0}} 
\ee
in fact, also extremizes (\ref{explicit}) for all values $U$ and
describes a static configuration of the D-string.  Here $r_0$ is the
position of the endpoint of the D-string along $r$ when $U=U_0$.  This
solution is position dependent and corresponds to a D-string stretched
between D3-branes extending along Melvin geometry.

In the case of constant non-commutativity, it was found that the
magnetic monopole becomes a dipole with a finite length proportional
to the non-commutativity parameter $\Delta$ and the expectation value
of the adjoint Higgs field $\Phi$ \cite{Hashimoto:1999zw,Gross:2000ss}
\be
L = 2 \pi \Delta^2 \Phi \ .
\ee
In the Melvin-Twist gauge theory where the non-commutativity parameter is non-constant, we find that the length of the dipole is position dependent.  Explicitly 
\be
L = r_0 \left( e^{2 \pi \Delta \Phi} - 1 \right) \ .
\ee
Therefore the length of the dipole is proportional to the local
magnitude of the non-commutativity parameter. For small values of
$\Delta \Phi$, the relation between the non-commutativity parameter
and the length of the dipole is exactly the same as in the case of
constant non-commutativity.  For arbitrary values of $\Delta \Phi$,
the magnetic monopoles acquire length in the radial direction.  This
suggests that the S-dual of this theory, where these monopoles become
the fundamental degrees of freedom, is an NCOS
\cite{Seiberg:2000ms,Gopakumar:2000na,Barbon:2000sg} with
non-vanishing commutation relations between the $t$ and $r$
coordinates.\footnote{Supergravity dual of NCOS with time dependent
non-commutativity parameter which arises as the S-dual of
\cite{Hashimoto:2002nr} is described in \cite{Cai:2002sv}.}

It is also straightforward to calculate the mass of the monopole from the DBI action.  We find
\be
M = \frac{2 \pi \Phi}{g_{YM}^2}  \ .
\ee
This mass is identical to the ordinary SYM monopole.  Interestingly,
this indicates that although the length of the monopole is position
dependent the mass is not and agrees with the constant
non-commutativity case.  It is also interesting that even though
supersymmetry is broken, these monopoles mimic BPS monopoles and can
therefore be moved freely through out the D3-brane world volume
without feeling any force.  These monopoles therefore serve as a
useful physical probe to measure the position dependence of the
non-commutativity parameter.

\section{Conclusion}

The purpose of this article was to explore physical aspects of a
non-commutative gauge theory with non-constant non-commutativity.  As
a particular example, we focused on the Melvin-Twist non-commutative
gauge theory which can be viewed as a special case of the gauge theory
whose action was constructed in \cite{Behr:2003qc}.  What makes the
model considered in this paper special is that it arises as the
decoupling limit of open strings ending on D3-branes extended along
the Melvin universe supported by the flux of an NSNS $B$-field
\cite{Hashimoto:2004pb}.  As such, we were able to utilize techniques
in string theory to extract physical information from this theory.
Specifically, we studied the supersymmetry, the Coulomb branch, and
the Prasad-Sommerfield magnetic monopoles, using the dual supergravity
formulation.

The matter content of the model under consideration is that of ${\cal
N} = 4$ SYM theory in four dimensions.  Although the free theory is
supersymmetric, interactions induced by the non-commutativity
necessarily break supersymmetry as translation invariance is broken by
the non-commutativity parameter. Interestingly, although supersymmetry
is broken in this model, we found that no potential is generated along
the Coulomb branch through quantum corrections in the large $N$ and
large 't Hooft coupling limit.  In light of this, we constructed
classical configurations of D-strings stretched between D3-branes
representing Prasad-Sommerfield monopoles.  Using the brane
configuration, we uncovered the interesting fact that although the
mass of the monopoles are the same as in commutative SYM theory, the
length is position dependent.  This reflects the fact that the
non-commutativity parameter is non-constant and is in line with the
findings of \cite{Hashimoto:1999zw} for the constant non-commutativity
case.  In contrast with the constant non-commutativity case, our model
is non-supersymmetric and it is quite remarkable that the monpoles of
this theory mimic BPS monopoles.

The conclusions regarding the monopoles were based entirely on the
dual supergravity formalism.  However, since the action of the field
theory under study is known, it should be possible to repeat the
analysis of the Coulomb branch potential perturbatively
\cite{Bertolami:2003nm,Robbins:2003ry} and also construct exact
solutions using the Nahm construction along the lines of
\cite{Gross:2000ss}.  It would be interesting to see if such analysis
reproduces the basic properties of the monpoles uncovered using the
dual supergravity description.  One could further consider issues such
as deformations of the moduli-space metric due to the
non-commutativity and it's effect on monopole scattering
\cite{Gibbons:1986df}.

Lastly, let us comment that the brane probe analysis of the
supergravity dual carried out in this paper can be repeated for other
supergravity duals of non-commutative gauge theories with non-constant
non-commutativity
\cite{Robbins:2003ry,Dolan:2002px,Lowe:2003qy,Hashimoto:2004pb}.  Some
of these models have a time-dependent non-commutativity parameter and
may give rise to interesting new physics.

\section*{Acknowledgements}

We would like to thank
D.~Chung  and
F.~Petriello
for useful discussions. This work was supported in part by the DOE
grant DE-FG02-95ER40896 and funds from the University of Wisconsin.

\bibliography{melvin}\bibliographystyle{utphys}

\end{document}